\documentclass[twocolumn]{autart}    

\usepackage[utf8]{inputenc}
\usepackage[hyperfootnotes=false]{hyperref}
\usepackage{comment}
\usepackage{algorithm}
\newtheorem{definition}{Definition}[section]
\usepackage{algpseudocode}

\usepackage{graphicx}
\let\labelindent\relax
\usepackage{enumitem}
\usepackage{amsmath,bm}
\usepackage[version=4]{mhchem}
\usepackage{siunitx}
\usepackage{longtable,tabularx}
\newcommand{\R}{\mathbb{R}}
\usepackage{graphics} 
\usepackage{float}
\usepackage{epsfig} 
\usepackage{times} 
\usepackage{amsmath} 
\usepackage{amssymb}  
\usepackage{mathtools}
\usepackage{color}
\usepackage[dvipsnames]{xcolor}
\usepackage{hyperref}
 \newtheorem{assumption}{Assumption}
  \newtheorem{lemma}{Lemma}
  \newtheorem{theorem}{Theorem}
 \newtheorem{corollary}{Corollary}
\newcounter {dagger} 
\newcommand{\MPar}[1]{\marginpar{$\ddagger$\thedagger}\stepcounter{dagger}}
\usepackage{bm}
\usepackage{times}
\usepackage{soul}
\usepackage{subcaption}
\newcommand{\matlab}{\mbox{\textsc{Matlab}\texttrademark}}
\newcommand{\bob}[1]{{\color{ForestGreen} #1}} 
\newcommand{\one}{\mathbf{1}}
\newcommand{\zero}{\mathbf{0}}

\definecolor{paleGreen}{rgb}{.3, .7, .3}
\definecolor{coolBlue}{rgb}{.3, .5, 1}
\definecolor{rosePink}{rgb}{.9, .5, .4}
\definecolor{ghost}{rgb}{.8, .8, .8}

\usepackage{graphicx}          

\begin{document}
\begin{frontmatter}

%

\title{Control-orientation, conservation of mass and model-based control of compressible fluid networks\thanksref{footnoteinfo}} 

\thanks[footnoteinfo]{This work was supported by Solar Turbines {Incorporated}.}

\author[First]{Sven Br{\"u}ggemann}
\author[Second]{Robert H. Moroto}
\author[First]{Robert R. Bitmead}

\address[First]{Mechanical \&\ Aerospace Engineering Department, University of California, San Diego, CA 92093-0411, USA, (e-mails: \{sbruegge, rbitmead\}@eng.ucsd.edu)}
\address[Second]{R. H. Moroto was formerly with Solar Turbines {Incorporated}, San Diego CA 92123, USA (e-mail: rhmoroto@gmail.com).}

\begin{abstract}
We study a gas network flow regulation control problem showing the closed-loop consequences of using interconnected component models, which have been designed to preserve a variant of mass flow conservation without the inclusion of algebraic constraints into the dynamics. These are candidate \textit{control-oriented} models because they are linear state-space systems. This leads to a study of mass conservation in flow models and the inheritance of conservation at the network level when present at each component. Conservation is expressed as a transfer function property at DC. This property then is shown to imply the existence of integrators and other DC structure of the network model, which has important consequences for the subsequent control design. An example based on an industrial system is used to explore the facility of moving from modeling to automated interconnection or components to model reduction to digital controller design and performance evaluation. Throughout, the focus is on the teasing out of control orientation in modeling. The example shows a strong connection between the modeling and the controller design.
\end{abstract}
\end{frontmatter}
\section{Introduction}
Our purpose is to analyze more fully network modeling and controller design based on the control-oriented models of \cite{SvenRobertBobTCST2021}, where standard fluid models, such as \cite{BennerGrundelChapter:2018}, which incorporate algebraic constraints associated with conservation of mass and pressure continuity at junctions, are replaced by linear state-space models adapted to effect these properties inherently, i.e. without the inclusion of explicit constraints. These unit models may be aggregated to yield more complicated network models, which in turn may be used for multi-input multi-output (MIMO) control design with standard tools. We put to the test the validity of whether these units models from \cite{SvenRobertBobTCST2021,sven_rob_bob_arXiv} truly are control-oriented and actually capture conservation and continuity when combined into full network models. This is examined in detail, notably to establish the presence of integrators or DC structure in the composite models. The subsequent controller design using these models is conducted and shown to reflect the requirements of conservation in the plant.

Our concern is to blend targeted but reusable control-oriented modeling methods with subsequent standard model-based control via the specific objective of designing controllers for compressible fluid regulation in a gas processing facility. The particular {example} process is the Gas Compressor Test Facility (GCTF) of Solar Turbines Incorporated, sponsor of this work. The corresponding model is developed and presented elsewhere \cite{SvenRobertBobTCST2021,sven_rob_bob_arXiv} and consists of linear state-space unit models interconnectable to yield a plant-wide or network composite model. The focus here is on two core aspects: the inheritance of conservation of mass from the units to the network despite the absence of algebraic constraints and the consequences of this for the network model in the light of controller design; and, MIMO control design using these approximate models and cognizant of the role of the conservation of mass in these designs. Throughout, the emphasis lies on the utility of the methodology for MIMO control design in gas processing facilities. 

In the first part of the body of this paper, we establish: explicit interconnection rules; mass conservation properties of the composite network model inferred from those of the components; consequences for the low-frequency dynamics of the network systems, such as the presence of integrators and differentiators, which have a strong bearing on regulator design \cite{mauricioLinCtrlBook2017,BekerHollotChaitTAC2001,MiddletonAutom1991}. The second part of this paper is devoted to controller design for a prototype control-oriented model from \cite{SvenRobertBobTCST2021} for the GCTF. We endeavor to come to grips with what it means for a model to be control-oriented by referring to philosophers of science and the specifics of the GCTF target example.

The gas facility control problem is characterized as follows:
 \begin{enumerate}[label=(\roman*)]
 \item A MIMO digital controller is sought to regulate the measured bulk pressures to nominal values in the face of disturbances.
 \item There is no intention to control turbulent or resonant acoustic modes in the system. Accordingly, the sampling rate for the digital controller is one sample per second or below and anti-aliasing filters with cut-off frequency 0.4 Hz are included into sensor channels.
\item The flow and pressure actuation is effected with control valves responding to the command signals, perhaps via a local controller, which is not part of our analysis.
 \item The loop is subject to disturbance flows and pressures, which are roughly constant at this sampling rate. 
 \item Because temperature sensing is slower than the sampling rate, we treat temperatures as known parameters in the models, as we do for compressor speed.
 \end{enumerate}
The development of our control design and analysis will be mindful of these objectives, particularly the emphasis on regulation of bulk pressure aspects with this sample rate. The loop model of the GCTF will provide the basis for the design study.

This paper is structured as follows. Section~\ref{sec:PDE} provides a brief treatment of models described by hyperbolic partial differential equations (PDEs) and their spatial discretization to yield linear lumped-parameter state-space models. To facilitate understanding, this is conducted for current and voltage in a linear electrical transmission line and analogously for the flow and pressure in our mildly nonlinear pipe model. This leads to an analysis of conservation of current in the line and conservation of mass flow in the pipe as consequences and artful generalizations of conservation of charge and mass. Section~\ref{sec:reson} compares the features captured by the discretized and linearized models: resonant modes, bulk behavior and conservation. Section~\ref{sec:SFG} presents the interconnection rules for the unit models and develops consequences for the transfer functions of the interconnected systems. Conservation of mass is explicitly defined for linear systems as a transfer function property in Section~\ref{sec:con}. Sections~\ref{sec:interc} and \ref{sec:integ} show that mass conservation of a network model is inherited from conservation of every component model and that this property affects the DC plant dynamics. For networks with flow input signals only, it leads to the presence of an integrator in the system. The latter sections of the paper, Sections~\ref{sec:models}-\ref{sec:MIMO-control} provide an example of MIMO LQG control design for the GCTF loop model, which possesses an integrator.

As we stated above, the objective is to trace through the control design possibilities engendered by the replacement of algebraic constraints, leading to differential algebraic equations (DAEs), or bond graph models by state-space or signal flow graph models, which preserve the mass conservation properties. An interesting feature is how the conservation properties appear in network models and in turn affect the control design using standard \matlab\ tools. Hopefully, we tease out some answers to the question of what it means to be control-oriented.

\section{PDE element models}\label{sec:PDE}
To help fix ideas, we present the parallel derivation of two lumped-parameter linear state-space models from discretization and (where needed) linearization of their constituent PDEs. The transmission line model is included because it is well known, linear and exhibits conservation of electrical charge. It is an adjunct to the main fluid flow model and its manifestation of conservation of mass.
\subsection{Transmission lines}
\begin{figure}[ht]
\begin{centering}
\includegraphics[width=\columnwidth]{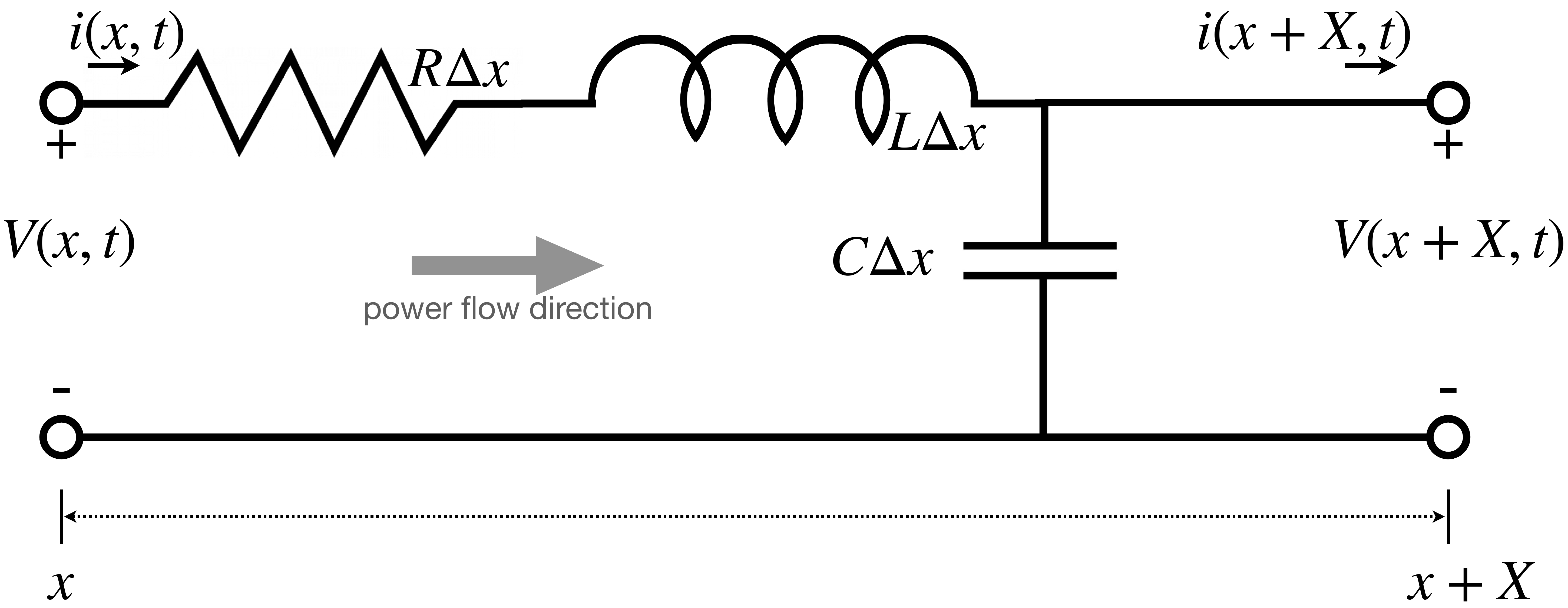}
\caption{Lumped-parameter transmission line model with zero shunt conductance.\label{fig:TxLine}}
\end{centering}
\end{figure}
A lumped-parameter transmission line segment of length $\Delta x$ with zero shunt conductance is depicted in Figure~\ref{fig:TxLine}. The partial differential equation (PDE) \textit{Telegraph Equations} describing this system are
\begin{align}
\frac{\partial i}{\partial x}(x,t)&=-C\frac{\partial v}{\partial t}(x,t),\label{eq:telei}\\
\frac{\partial v}{\partial x}(x,t)&=-Ri(x,t)-L\frac{\partial i}{\partial t}(x,t).\label{eq:telev}
\end{align}
Direct calculations yield
\begin{align}
\frac{\partial^2 i}{\partial x^2}&=RC\frac{\partial i}{\partial t}+LC\frac{\partial^2 i}{\partial t^2},\label{eq:wavei}\\
\frac{\partial^2 v}{\partial x^2}&=RC\frac{\partial v}{\partial t}+LC\frac{\partial^2 v}{\partial t^2},\label{eq:wavev}\\
\frac{\partial(iv)}{\partial x}
&=-Ri^2-\frac{\partial\left[\frac{1}{2}Li^2\right]}{\partial t}-\frac{\partial\left[\frac{1}{2}Cv^2\right]}{\partial t}.\label{eq:disscct}
\end{align}
Equations \eqref{eq:wavei} and \eqref{eq:wavev} are damped wave equations and \eqref{eq:disscct} captures the propagation, dissipation and storage of power along the line. The properties of the solutions, including energy and mass flow, are determined by the boundary conditions.

\subsection{One-dimensional pipe flow}
\begin{figure}[ht]
\begin{centering}
\includegraphics[width=\columnwidth]{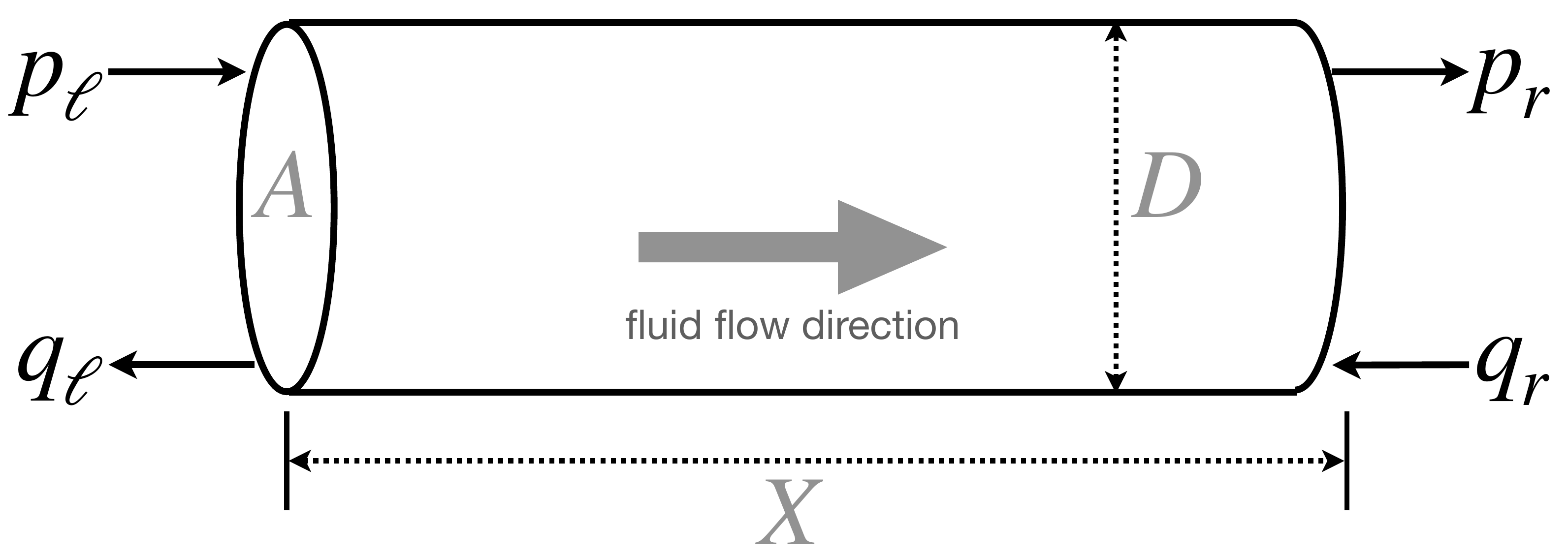}
\caption{Pipe diagram defining: flow and signal directions; length, diameter, and cross-sectional area.\label{fig:pipe}}
\end{centering}
\end{figure}
One-dimensional isothermal compressible gas flow in horizontal pipes may be described by the following partial differential equations, which combine: the Euler equations, Ideal Gas Equation, and  the assumption that the transport velocity is significantly lower than the speed of sound, $c=\sqrt{R_sT_0z_0}$. (See \cite{CengelCimbala:2013,BennerGrundelChapter:2018}.)
\begin{align}
\frac{\partial\check p}{\partial t}&=-\frac{c^2}{A}\frac{\partial\check q}{\partial x},\label{eq:qeqn}\\
\frac{\partial\check q}{\partial t}&=-A\frac{\partial\check  p}{\partial x}-\lambda\frac{c^2}{2DA}\frac{\check q|\check q|}{\check p}.\label{eq:peqn}
\end{align}
The solution of these equations defines pressure, $\check p(x,t)$, and mass flow, $\check q(x,t)$, as functions of space, $x\in[0,X]$ with length $X$, and time, $t\in(0,\infty)$.
Define the nominal values of pressure and flow to be $\bar p(x,t)$ and $\bar q(x,t)$. Further, define variations about these nominal values:
\begin{align*}
p(x,t)=\check p(x,t)-\bar p(x,t),\hskip 5mm q(x,t)=\check q(x,t)-\bar q(x,t).
\end{align*}
Then the linearization of the pipe flow PDEs, assuming flow $\check q$ is positive, is
\begin{align}
\frac{\partial p}{\partial t}&=-\frac{c^2}{A}\frac{\partial q}{\partial x},\label{eq:ppde}\\
\frac{\partial q}{\partial t}&=-A\frac{\partial p}{\partial x}-\frac{\lambda c^2}{DA}\frac{\bar q}{\bar p}q+\frac{\lambda c^2}{2DA}\frac{\bar q^2}{\bar p^2}p.\label{eq:qpde}
\end{align}
In turn, this yields damped wave equations for $q$ and for $p$, \textit{cf.} \eqref{eq:wavei}, \eqref{eq:wavev}.
\begin{align}
\frac{\partial^2 q}{\partial x^2}-\frac{\lambda c^2}{2DA^2}\frac{\bar q^2}{\bar p^2}\frac{\partial q}{\partial x}&=\frac{1}{c^2}\frac{\partial^2 q}{\partial t^2}+\frac{\lambda}{DA}\frac{\bar q}{\bar p}\frac{\partial q}{\partial t},\label{eq:waveq}\\
\frac{\partial^2 p}{\partial x^2}-\frac{\lambda c^2}{2DA^2}\frac{\bar q^2}{\bar p^2}\frac{\partial p}{\partial x}&=\frac{1}{c^2}\frac{\partial^2 p}{\partial t^2}+\frac{\lambda}{DA}\frac{\bar q}{\bar p}\frac{\partial p}{\partial t}.\label{eq:wavep}
\end{align}

\subsection{Spatial discretization and state-space models}
Regarding the transmission line as a two-port network, we recognize that circuit variables, $i$ and $v$, cannot both be independently prescribed at a single port. Accordingly, we treat $v_\ell(t)\stackrel{\triangle}{=}V(x,t)$ and $i_r(t)\stackrel{\triangle}{=}i(x+X,t)$ as the input variables and $v_r(t)\stackrel{\triangle}{=}V(x+X,t)$ and $i_\ell(t)\stackrel{\triangle}{=}i(x,t)$ as output or response variables. This choice of $(v_\ell,i_r)$ as the free input signals coincides with the (mixed Dirichlet, Neumann) boundary condition specification required for the hyperbolic damped wave equation, \eqref{eq:wavei} or \eqref{eq:wavev}. This is explained in more detail for the pipe model in \cite{SvenRobertBobTCST2021}. Spatial discretization of (\ref{eq:telei}-\ref{eq:telev}) yields a linear state-space description.
\begin{align}
\begin{bmatrix}\dot v_r\\\dot i_\ell\end{bmatrix}&=\begin{bmatrix}0&\frac{1}{CX}\\-\frac{1}{LX}&-\frac{R}{L}\end{bmatrix}\begin{bmatrix}v_r\\i_\ell\end{bmatrix}
+\begin{bmatrix}0&-\frac{1}{CX}\\\frac{1}{LX}&0\end{bmatrix}\begin{bmatrix}v_\ell\\i_r\end{bmatrix}.\label{eq:txss}
\end{align}

Similarly, since for the pipe model pressure and flow behave analogously to voltage and current in the transmission line and cannot be separately prescribed functions of time at the same point in space and \eqref{eq:waveq} and \eqref{eq:wavep} are also a damped wave equations, define
\begin{align*}
p_\ell&=p(x,t),&p_r&=p(x+X,t),\\
q_\ell&=q(x,t),&q_r&=q(x+X,t).
\end{align*}
The discretized and linearized state-space equations are\footnote{The friction correction term to $\frac{A}{X}$, $\frac{\lambda c^2}{2DA}\frac{\bar q^2}{\bar p^2}$ may be taken into account in the system or input matrix and appears because $\bar p_r=\left(1-\frac{\lambda c^2X}{2DA^2}\frac{\bar q_r^2}{\bar p_\ell^2}\right)\bar p_\ell.$ It is a small term compared to $\frac{A}{X}$.}
\begin{align}
\begin{bmatrix}\dot p_r\\\dot q_\ell\end{bmatrix}&=\begin{bmatrix}0&\frac{c^2}{AX}\\-\frac{A}{X}&-\frac{\lambda c^2}{DA}\frac{\bar q_r}{\bar p_\ell}\end{bmatrix}
\begin{bmatrix}p_r\\q_\ell\end{bmatrix}\nonumber\\&\hskip 20mm+\begin{bmatrix}0&-\frac{c^2}{AX}\\\frac{A}{X}+\frac{\lambda c^2}{2DA}\frac{\bar q_r^2}{\bar p_\ell^2}&0\end{bmatrix}
\begin{bmatrix}p_\ell\\q_r\end{bmatrix}.\label{eq:pipess}
\end{align}

\section{Model properties for pipe and transmission line}\label{sec:reson}
\begin{figure}[ht]
\begin{centering}
\includegraphics[width=\columnwidth]{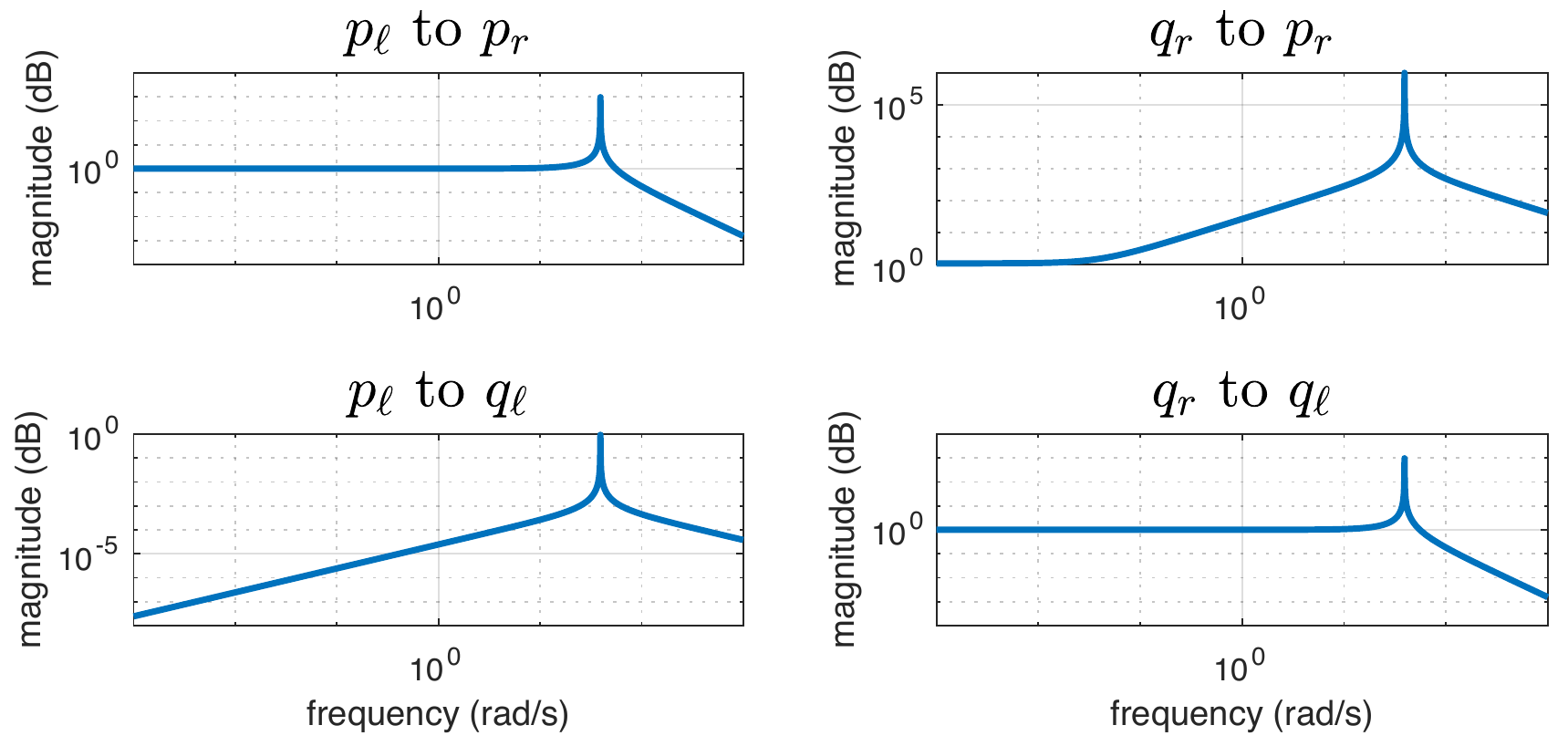}
\caption{Pipe Bode magnitude plots: $(p_\ell,q_r)$ to $(p_r,q_\ell)$.\label{fig:pipeBode}}
\end{centering}
\end{figure}

The damped wave equations, (\ref{eq:wavei}-\ref{eq:wavev}) for the transmission line and (\ref{eq:waveq}-\ref{eq:wavep}) for the pipe, and the spatially discretized and linearized state-space variants, \eqref{eq:txss} and \eqref{eq:pipess} respectively, exhibit wave resonance and damping due to resistance or friction. For the transmission line model, the eigenvalues are at $-\frac{R}{2L}\pm\sqrt{\frac{R^2}{4L^2}-\frac{1}{(LCX^2)}}$, while for the pipe they are at $-\frac{\lambda c^2}{2DA}\frac{\bar q}{\bar p}\pm\sqrt{\frac{(\lambda c^2)^2}{4(DA)^2}\frac{\bar q^2}{\bar p^2}-\frac{c^2}{X^2}}$. Since the $R$ and $\lambda$ terms are small, the surds yield imaginary numbers, whose physical interpretation as oscillations is immediate, even though one refers to transverse electromagnetic propagation and the other to acoustic compression waves. The negative real parts, likewise, are simply interpreted. Figure~\ref{fig:pipeBode} shows the four frequency response magnitudes for a representative 10m steel gas pipe.

The acoustic resonant mode, here at 6.25Hz, falls well above the Nyquist sampling frequency of 0.25-0.5Hz. The DC gains of \eqref{eq:txss} and \eqref{eq:pipess} are given symbolically by
\begin{align}
\hskip -3mm\text{G}_\eqref{eq:txss}=\begin{bmatrix}1&-RX\\0&1\end{bmatrix},\;
\text{G}_\eqref{eq:pipess}=\begin{bmatrix}1+\frac{\lambda c^2X}{2DA^2}\frac{\bar q^2}{\bar p^2}&-\frac{X\lambda c^2}{DA^2}\frac{\bar q}{\bar p}\\0&1\end{bmatrix}.
\label{eq:dcgain}
\end{align}

\section{Signal Flow Graph models}\label{sec:SFG}
\subsection{Philosophy of signal flow graph modeling}
The input-output representations implicit in \eqref{eq:txss} and \eqref{eq:pipess} conform to viewing the transmission line or pipe segments as signal-flow graphs with directed connections. This is in keeping with their eventual application as part of a feedback control system. These ideas date back to Claude Shannon \cite{ShannonCollected1993} and Samuel Mason \cite{Mason1953} and may be contrasted with the undirected graph methods of Jan Willems' \textit{Behavioral Theory} \cite{WillemsCSM2007,WillemsCASM2010}, including electrical and mechanical analogs \cite{WillemsElectricalCDC2013,WillemsMechanicalCDC2013}, and directed DAE methods associated with \textit{Bond Graphs} \cite{BennerGrundelChapter:2018,GawthropBevanCSM2007}. The critical aspect of signal-flow graph representations is their conformity with control design tools, such as those in \matlab. The price of directed graph methods is that they entail an implied (and somewhat arbitrary) causality of the interconnection, which in electrical networks would imply buffering at junctions. However, as remarked above, there is direct relation to the requisite boundary conditions and this choice of signals and state.

As Laudan \cite{Laudan:1996} states ``the aim of science is to secure theories with a high problem-solving effectiveness.'' And this approach to lumped-parameter, linearized, spatially discretized, signal flow graph modeling operates with this recognition of the convenient fiction underpinning all modeling but reflective of the utility of the approach. Indeed, Harold Black's original analysis of the analog negative feedback amplifier \cite{BlackHarold1934} adopts this approach. Newcomb \cite{NewcombBook1966} and Anderson \&\ Vongpanitlerd \cite{AndersonVongpanitlerd:73} identify ports and driving-point impedances for this purpose. Enough name dropping. For us, it permits access to \matlab's design and analysis tools and network modeling from component subsystems; very utile indeed.

\subsection{Directed pipe connections and `ports'}
The signal flow graph models derived above have directions associated with each signal. Thus, $p_\ell$ and $q_r$ are input signals, indicating that they are specified from outside the pipe, and $p_r$ and $q_\ell$ are output signals, meaning that they are determined by the pipe system and the input signals. The spatially localized connection sites, however, possess one input signal and one output signal, namely $(p_\ell,q_\ell)$ at the left end and $(p_r,q_r)$ at the right. We further appropriate circuit terminology and identify two distinct location types, which we term \textit{ports}. As in \cite{BennerGrundelChapter:2018}, every element in our interconnected system presents pipe signal interfaces to other elements and to the outside world.

\begin{definition}[Ports]
\begin{description}
\item
\item[$\boldsymbol p$-port]  of a component possesses two signals: an input pressure signal $p_\ell$ and an output flow signal $q_\ell$.
\item[$\boldsymbol q$-port] of a component possesses two signals: an input flow signal $q_r$ and an output pressure signal $p_r$.
\end{description}
\end{definition}

Internal series connection of two components, 1 and 2, will involve the cascading of signals $p^2_\ell=p^1_r$ and $q^1_r=q^2_\ell$ at the junction point. This describes a $p$-port to $q$-port connection. Likewise, connection to the outside of the network must respect the type and causality of the signals. These rules are specified below.

\textbf{Interconnection Rules}

\begin{enumerate}[label=\Roman*.]
\item 
Connections are permitted only between:
\begin{enumerate}[label=\roman*.]
\item a $p$-port and a $q$-port, or
\item a $p$-port and an external pressure source/input signal plus an external flow sink/output signal, or
\item a $q$-port and an external flow source/input signal plus an external pressure sink/output signal.
\end{enumerate}
\item Pressure input signals must connect to pressure output signals and flow input signals must connect to flow output signals. 
\item Connection of one variable of a port requires connection of the other.
\item All ports must be connected and algebraic loops avoided.
\end{enumerate}
These rules conform to the connections examined in \cite{SvenRobertBobTCST2021} to formulate the systematic interconnection of state-space models. Each component model possesses input and output signals and, in Proposition~3 \cite{SvenRobertBobTCST2021}, it is shown how a (possibly non-minimal) state-space realization of the interconnection of gas system elements can be directly constructed with the above rules. This construction replaces and extends the graph-theoretic DAE methods of \cite{BennerGrundelChapter:2018} and yields a new input-output transfer function satisfying Mason's Gain Formula, \cite[Proposition~4]{SvenRobertBobTCST2021}. 

Per \cite{SvenRobertBobTCST2021,sven_rob_bob_arXiv}, the fluid-flow network model commences with an aggregate (direct sum) model of all elements in state-space form,
\begin{align}\label{eq:sys_network}
\dot{ x}&=A x+B w,\;\;\; y=C x+D w,
\end{align}
with $x\in\R^{n_x},w\in\R^{n_w}$ and $y\in\R^{n_y}$.
Interconnections and external sources $u\in\R^{n_u}$ {and sinks $z\in\R^{n_z}$} are described by 
\begin{align}\label{eq:FG_network}
 w = F y+G u,\;\;\; z = Hx+J u,
\end{align}
with {structured matrices} $[F,G,H,J]$ {with 0-1 elements}.

The Interconnection Rules allow us to group outputs by their type, pressure or flow, and connection, internal or external. Denote the (row-organized) collection of input and output signals as 
\begin{align}\label{eq:y=zy}
y=\begin{bmatrix}
z_p\\ z_q\\ \tilde y_p\\ \tilde y_q
\end{bmatrix},\hskip 5mm
w=\begin{bmatrix}
 u_{p} \\ u_{q}\\ \tilde w_{p}\\\tilde w_{q}
\end{bmatrix},
\end{align}
where, for the connected network: 
$z_p\in\R^{n_{zp}}$ and $z_q\in\R^{n_{zq}}$ are external pressure and mass flow output signals;
$ u_p\in\R^{n_{uq}}$ and $ u_q\in\R^{n_{uq}}$ are external pressure and mass flow input sources;
$\tilde y_p\in\R^{n_{\tilde yp}}$ and $\tilde y_q\in\R^{n_{\tilde yq}}$ are internal pressure and mass flow output signals;
$\tilde w_p\in\R^{n_{\tilde wp}}$ and $\tilde w_q\in\R^{n_{\tilde wq}}$ are internal pressure and mass flow input signals.

\begin{lemma}\label{lm:portcount}
The connection rules imply the following.
\begin{enumerate}[label=(\roman*)]
\item $\tilde w_q=\tilde y_q$ whence $n_{\tilde wq}=n_{\tilde yq}$.
\item $\tilde w_p=\tilde y_p$ whence $n_{\tilde wp}=n_{\tilde yp}$.
\item The number of external pressure input signals equals the number of external flow output signals: $n_{up}=n_{zq}$. 
\item The number of external pressure output signals equals the number of external flow input signals: $n_{zp}=n_{uq}$.
\end{enumerate}
\end{lemma}

We denote the unconnected network by the transfer function matrix, suppressing the $s$-dependence,
\begin{align}\label{eq:unco}
\begin{bmatrix}
Z_p\\Z_q\\\tilde Y_p\\\tilde Y_q
\end{bmatrix}
&=\begin{bmatrix}
T_{zp,up}& T_{zp,uq}&T_{zp,\tilde w p}& T_{zp,\tilde w q}\\
T_{zq,up}&T_{zq,uq}& T_{zq,\tilde w p}& T_{zq,\tilde w q}\\
T_{\tilde yp,up}&T_{\tilde yp,uq}& T_{\tilde yp,\tilde w p}& T_{\tilde yp,\tilde w q}\\
T_{\tilde yq,up}&T_{\tilde yq,uq}&T_{\tilde yq,\tilde w p}&T_{\tilde yq,\tilde w q}
\end{bmatrix}\begin{bmatrix}U_p\\U_q\\\tilde W_p\\\tilde W_q\end{bmatrix},
\end{align}
and the connected transfer function matrix, formed by connecting $\tilde w_p=\tilde y_p$ and $\tilde w_q=\tilde y_q,$ by
\begin{align}\label{eq:fsys}
\begin{bmatrix}Z_p(s)\\Z_q(s)\end{bmatrix}=\begin{bmatrix}T_{pp}(s)&T_{pq}(s)\\T_{qp}(s)&T_{qq}(s)\end{bmatrix}\begin{bmatrix}U_p(s)\\U_q(s)\end{bmatrix}.
\end{align}

\section{Conservation properties of state-space models}\label{sec:con}
In \eqref{eq:dcgain} above, the algebraic unity DC-gain in $[\text{G}_\eqref{eq:pipess}]_{2,2}$ together with zero DC-gain in $[\text{G}_\eqref{eq:pipess}]_{2,1}$ suggest conservation rules at play. However, while charge and mass are conserved, instantaneous current and mass flow need not be because of the circuit capacitance and the gas compressibility. Integrating \eqref{eq:telei} and \eqref{eq:ppde} spatially from left to right, we see that
\begin{align}
i_r&=i_\ell -C\int_\ell^r{\frac{\partial v}{\partial t}\,dx},\nonumber\\
q_r&=q_\ell-\frac{A}{c^2}\int_\ell^r{\frac{\partial p}{\partial t}\,dx}.\label{eq:intgrl}
\end{align}
That is, the deficit in current conservation is attributed to the time derivative of the voltage along the line, and that in mass flow conservation to the pressure time-variation along the pipe. Conservation in flow variables occurs when these time-derivatives are zero. That is, at steady state. From this point on, we drop the parallel references to the transmission line problem and concentrate solely on fluid flow systems. 

For the connected fluid network \eqref{eq:fsys}, and with a slight abuse of language -- \textit{mass} for \textit{mass flow} -- we make the following definition.
\begin{definition}\label{def:com}
For a linear fluid system with transfer function matrix as in \eqref{eq:fsys}, we say \textit{the system conserves mass} if
\begin{align*}
\lim_{s\to 0}{\mathbf 1_{n_{zq}}T_{qq}(s)}=\mathbf 1_{n_{uq}}\text{  and  }\lim_{s\to 0}T_{qp}(s)=\mathbf 0,
\end{align*}
where $\mathbf 1_n$ is a row vector of ones in $\mathbb R_n$ and $\mathbf 0$ is the zero matrix of appropriate dimensions. 
\end{definition}
The upshot from this definition is that, if the system \eqref{eq:fsys} possesses a steady-state (and Definition~\ref{def:com} does not require stability), then 
\begin{align*}\lim_{t\to\infty}{\mathbf 1_{n_{zq}}z_q(t)}=\lim_{t\to\infty}{\mathbf 1_{n_{uq}}u_q(t)},\end{align*}
for the associated steady-state time-domain flow variables.
As we see from \eqref{eq:dcgain}, our linearized models of pipe elements satisfy conservation of mass flow. Other system elements, such as branches, joints, compressors, valves, heat exchangers, tanks, manifolds derived in \cite{SvenRobertBobTCST2021,sven_rob_bob_arXiv}, also are seen to possess this property.

Lemma~\ref{lm:portcount} has the following consequence.
\begin{corollary}\label{cor:shape}
If the connection rules apply, then the matrices in \eqref{eq:fsys} have the following dimensions
\begin{align*}
\dim T_{pp}&=n_{zp}\times n_{up}=n_{uq}\times n_{up},\\
\dim T_{pq}&=n_{zp}\times n_{uq}=n_{uq}\times n_{uq},\\
\dim T_{qp}&=n_{zq}\times n_{up}=n_{up}\times n_{up},\\
\dim T_{qq}&=n_{zq}\times n_{uq}=n_{up}\times n_{uq}.
\end{align*}
\end{corollary}

\section{Interconnection of conservative gas flow elements}\label{sec:interc}
\begin{assumption}\label{ass:dim}
For each network component, conservation of mass is satisfied and the interconnection rules between components are satisfied. 
\end{assumption}
This assumption is satisfied for all component models from \cite{SvenRobertBobTCST2021,sven_rob_bob_arXiv}.

We have the following property of interconnected networks.
\begin{theorem}[Mass-conserving network]\label{thm:cof_network}
Subject to Assumption~\ref{ass:dim}, the network also satisfies conservation of mass.
\end{theorem}
\textit{Proof:} We have $y_q^\top=\begin{bmatrix}
z_q^\top & \tilde y_q^\top
\end{bmatrix}\in\R^{n_{yq}},$
with $n_{yq}=n_{zq}+n_{\tilde yq}$, and conformably for $w_q$. Allowing signals to reach steady state (which we denote by subscript $(\cdot)_{,ss}$), by hypothesis of mass-conserving elements we have,
\vspace{-10pt}
\begin{align*}
\one_{n_{yq}} y_{q,ss}&=\one_{n_{wq}} w_{q,ss},\\
\begin{bmatrix}
\one_{n_{zq}}&\one_{n_{\tilde yq}}
\end{bmatrix}\begin{bmatrix}
z_{q,ss} \\ \tilde y_{q,ss}
\end{bmatrix}&=\begin{bmatrix}
\one_{n_{uq}}&\one_{n_{\tilde wq}}
\end{bmatrix} \begin{bmatrix}
u_{q,ss} \\ \tilde w_{q,ss}
\end{bmatrix},
\end{align*}
From Lemma~\ref{lm:portcount}, $\tilde y_q=\tilde w_q$, so 
\vspace{-5pt}
\begin{align*}
\begin{bmatrix}
\one_{n_{zq}} & \one_{n_{\tilde yq}}
\end{bmatrix} \begin{bmatrix}
z_{q,ss} \\ \tilde y_{q,ss}
\end{bmatrix}&=\begin{bmatrix}
\one_{n_{uq}} & \one_{n_{\tilde yq}}
\end{bmatrix} \begin{bmatrix}
u_{q,ss} \\ \tilde y_{q,ss}
\end{bmatrix},\\
\one_{n_{\tilde yq}}
z_{q,ss}&=\one_{n_{uq}} u_{q,ss}.
\end{align*}
This is conservation of mass of the connected system.
\hfill$\square$

\section{Integrators in mass-conserving networks}\label{sec:integ}
The presence of integrators in transfer functions is important for the design of feedback regulator control systems. We next establish that mass-conserving fluid networks implicitly possess integrators in the mass-flow to pressure path. Since the pressure signal is most easily measured, admitting rapid and accurate sampling versus, say, the mass flow or temperature signals, pressure provides the most common signal for disturbance rejection feedback, target control specification and output sensing. By contrast, mass flow is frequently the dominant input variable effected by valves. These integrators, in the fluid system and in the network models here, are central to the design of effective feedback controls and hence the control relevance of these models and methodology for their construction.

The fluid dynamics Continuity Equation \eqref{eq:ppde} applied to a single pipe exhibits the integration property.
\begin{align*}
\frac{\partial p}{\partial t}&=-\frac{c^2}{A}\frac{\partial q}{\partial x}.
\end{align*}
This may be spatially discretized and then integrated with respect to time to yield
\begin{align}
p(\bar x,t)-p(\bar x,t_0)&=-\frac{c^2}{AL}\int^t_{t_0}{[q_r(\tau)-q_\ell(\tau)]\,d\tau}.\label{eq:ctyint}
\end{align}
Here from the Mean Value Theorem, $\bar x$ is a point inside the $(x,x+L)$ interval. In physical terms, \eqref{eq:ctyint} reflects the property that a steady-state mismatch between the flows into and out from the pipe results in an unbounded change in pressure.

This analysis may be extended to the network behavior.
\begin{theorem}\label{thm:unbound}
For a fluid network satisfying Assumption~\ref{ass:dim} and possessing pressure input signals, the transfer function $T_{qp}(s)$ in \eqref{eq:fsys} possesses a blocking zero at $s=0$. So, a steady-state difference between inflows and outflows can only be achieved by unbounded pressure inputs.
\end{theorem}
\textit{Proof:}
From Definition~\ref{def:com}, conservation of mass entails $T_{qp}(0)=\zero$; hence the first part of the statement. Now, from the Laplace transform Final Value Theorem, the steady-state difference between net flow out and net flow in is given by
\begin{align}
d&=\lim_{s\to 0}{s\left[\one_{n_{zq}}Z_q(s)-\one_{n_{uq}}U_q(s)\right]}\nonumber\\
&=\lim_{s\to 0}{s\one_{n_{zp}}T_{qp}(s)},\label{eq:diffo}
\end{align}
where we have used
\begin{align*}
Z_q(s)&=T_{qp}(s)U_p(s)+T_{qq}(s)U_q(s),
\end{align*}
from \eqref{eq:fsys} rewritten as
\begin{align}
T_{qp}(s)U_p(s)&=Z_q(s)-T_{qq}(s)U_q(s),\label{eq:up1}
\end{align}
and first equality for conservation of mass from Definition~\ref{def:com}. Now if $d$ in \eqref{eq:diffo} is non-zero, then
\begin{align*}
\one_{n_{zp}}T_{qp}(s)U_p(s)&=\frac{d}{s}+\text{terms with no poles at $s=0$}.
\end{align*}
Since $T_{qp}(0)=0$, if $d\neq 0$ then $U_p(s)$ must have a double pole at $s=0$.
\hfill$\square$

For connected networks with only flow input signals and only pressure output signals, that is from Lemma~\ref{lm:portcount}, $n_{zq}=n_{up}=0$ and $n_{zp}=n_{uq}>0$, we make the following assumption.
\begin{assumption}\label{ass:connected}
The unconnected network satisfies $n_{\tilde wq}>0$ and has $T_{zp,\tilde wq}(s)$ full rank at $s=0$.
\end{assumption}
This assumption is evidently an observability or, as we shall see, detectability condition. It ensures that the externally measured pressure signals suffice to reveal static mismatches in internal flows. Inspecting the steady-state gains, \eqref{eq:dcgain}, of the fundamental pipe equation, \eqref{eq:pipess}, or its transmission line precursor, \eqref{eq:txss}, the satisfaction of Assumption~\ref{ass:connected} is assured provided that the pipe friction is non-zero or the line resistance is non-zero. It is a testable condition for the network and is evidently satisfied by the pipe element in the upper right figure of Figure~\ref{fig:pipeBode}.

With this new assumption, we can now state the central result.
\begin{theorem}[Network with integrator]\label{thm:integrator1}
Consider the fluid network with external input signals, which are mass flows alone and satisfying Assumptions~\ref{ass:dim} and \ref{ass:connected}. Then the output pressure signals contain a term proportional to the integral of $\one_{n_{uq}}u_q(t)$, the mass flow mismatch into the network.
\end{theorem}
\textit{Proof:}
We have $n_{zq}=n_{up}=0$ and $n_{zp}=n_{uq}\neq 0$.
From \eqref{eq:unco}, identify the non-zero-dimension submatrices of the unconnected system's transfer function and extract the component output flow signals.
\begin{align}\label{eq:tilde}
\hskip -3mm\tilde Y_q(s)&=T_{\tilde yq,uq}U_q(s)+T_{\tilde yq,\tilde wp}\tilde W_p(s)+T_{\tilde yq,\tilde wq}\tilde W_q(s).
\end{align}
From conservation of mass of the individual network components and $n_{zq}=0$, we have that $n_{yq}=n_{\tilde yq}$ and
\begin{align*}
\one_{n_{\tilde yq}} \begin{bmatrix}
T_{\tilde yq,uq}(0) & T_{\tilde yq,\tilde w q}(0)
\end{bmatrix}=\begin{bmatrix}
\one_{n_{uq}}&\one_{n_{\tilde wq}}
\end{bmatrix},
\end{align*}
or {since} $n_{\tilde yq}=n_{\tilde wq}$,
\begin{align*}
\one_{n_{\tilde yq}}
T_{\tilde yq,\tilde wq}(0)=\one_{n_{\tilde yq}}.
\end{align*}
That is $T_{\tilde yq,\tilde wq}(0)$ has an eigenvalue at $1$ with left eigenvector $\one_{n_{\tilde yq}}$. Additionally, we have
\begin{align*}
\lim_{s\to 0}{T_{\tilde yq,\tilde wp}}&=0.
\end{align*}

Returning to \eqref{eq:tilde} and substituting for the connection $\tilde W_q(s)=\tilde Y_q(s)$,
\begin{align*}
\lim_{s\to 0}{\one_{n_{yq}}\tilde Y_q}&=\lim_{s\to 0}{\one_{n_{uq}}U_q}+\lim_{s\to 0}{\one_{n_{yq}}\tilde Y_q}.
\end{align*}
Thus, to first order in $s$,
\begin{align*}
s\one_{n_{yq}}\tilde Y_q(s)&=\one_{n_{uq}}U_q(s).
\end{align*}
That is, the internal flow signal vector, $\tilde y_q(t),$ contains the integral of $\one_{n_{uq}}u_q(t)$, the integral of the network flow mismatch.

The (output) pressure signals satisfy, after substituting for the connection $\tilde W_p(s)=\tilde Z_p(s)$,
\begin{align*}
Z_p(s)&=(I-T_{zp,\tilde wp})^{-1}\left[T_{zp,uq}U_q(s)+T_{zp,\tilde wq}\tilde Y_q(s)\right].
\end{align*}
Assumption~\ref{ass:connected} ensures that integral of $\one_{n_{uq}}u_q(t)$ appears in the output pressure.
\hfill$\square$

The presence of integrators in the pressure measurements of gas flow networks with only external flow signals appearing as inputs, is a manifestation of the physical property that, if the net mass flow into a network is unbalanced, i.e. sums to a non-zero number, in steady state, then this is reflected in the output pressures becoming unbounded. Of course, this is a manifestation of conservation of mass. But, from a control systems perspective, it is important to demonstrate that this is (i) a property captured by the approximate models and their aggregation into network models, and (ii) manifested through the appearance of an integrator, whose role in feedback control design is to ensure mass flow balance without the imposition of a constraint to achieve this.

\section{Control-oriented component and network models}\label{sec:models}
Linearized MIMO state-space component models have been developed in \cite{SvenRobertBobTCST2021,sven_rob_bob_arXiv} based on extensions of the pipe models \eqref{eq:pipess} to include branch and join combinations, tanks, valves, manifolds, etc. As explained earlier, these models themselves replace DAE constructions from \cite{BennerGrundelChapter:2018} for distribution systems and make the conceptual leap from physical models to directed signal flow graphs. Compressor models have been adapted from \cite{EgelandGravdahl2002}; see \cite{sven_rob_bob_arXiv}. These component models each possess the following features.
\begin{enumerate}[label=(\roman*)]
\item They are finite-dimensional linear state-space models with flows, pressures and temperatures as signals and states. Isothermal variants are possible with gas temperatures appearing as parameters.
\item They conserve mass per Definition~\ref{def:com}. 
\item They satisfy Assumption~\ref{ass:connected}.
\item They are continuous-time and capture both bulk modes and resonant behaviors.
\end{enumerate}
The component models may be connected using the \matlab\ methods described in \cite{SvenRobertBobTCST2021} or \matlab's \texttt{connect} function \cite{sven_rob_bob_arXiv}, which comply with the Interconnection Rules of Section~\ref{sec:SFG} and the analyses above. Therefore, the predicates of Theorems~\ref{thm:cof_network}, \ref{thm:unbound} and \ref{thm:integrator1} abide.

But what makes them \textit{control-oriented?} This requires the specification of a control problem and the application of the network models for control design. This gas processing facility regulation problem was presented in the Introduction and has guided the analysis so far. It is worth stating that the price paid for control-orientation is a departure from verisimilitude in high-frequency fluid flow modeling and adoption of directed graph models, which ignore back flows and pressures. The benefit is simplicity and amenability to the tools of controller design. Ultimately, however, it is the utility of these models for the construction of MIMO feedback control, which is the only meaningful metric of performance or acceptability.

\section{MIMO control loop}\label{sec:MIMO-control}
\begin{figure}[ht!]
    \centering
   \includegraphics[width=\columnwidth]{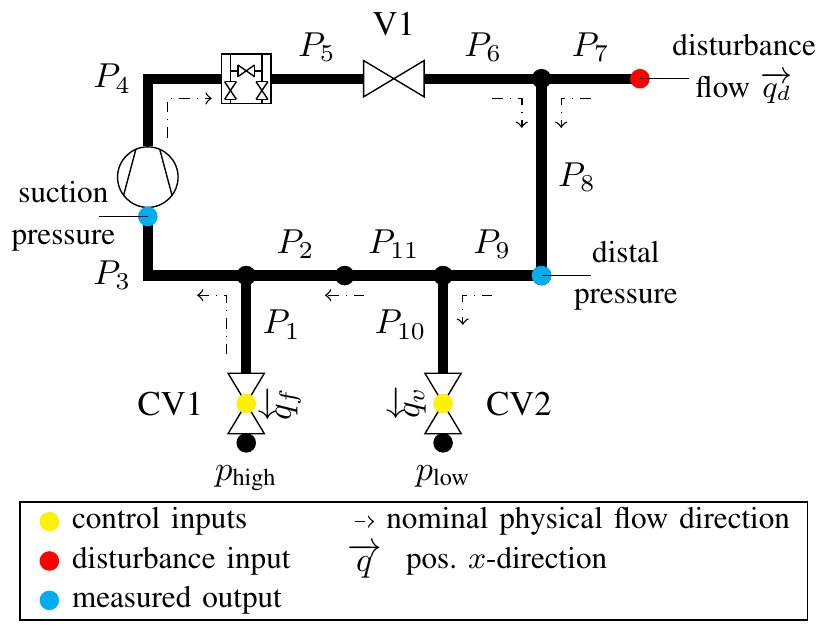}
  \caption{Pipe network with compressor, valves and valve manifold. 
  \label{fig:loop}}
  \end{figure}

The control objective is to regulate the pressures in the gas loop depicted in Figure~\ref{fig:loop}. This is a representation of the Solar Turbines Incorporated GCTF in San Diego, California, which is used to evaluate compressor performance in operation. The GCTF is: well instrumented; of a scale appropriate to gas handling facilities; equipped with safety control systems over and above regulation and process control objectives, which are presently handled by local SISO PI controllers and which we seek eventually to replace via the systematic introduction of MIMO control based on our aggregated models. At this stage, proof of concept MIMO design is the objective prior to in-situ testing of regulation performance.
  
 The general problem statement in the Introduction is qualified as follows.
  \begin{itemize}
  \item A two-input/two-output digital MIMO controller is sought to regulate the two measured pressures: compressor suction pressure, $p_\text{suc},$ and distal pressure, $p_\text{dstl}$, shown as blue dots in Figure~\ref{fig:loop}. There is particular emphasis on managing $p_\text{suc}$, the inlet pressure to the compressor.
 \item The two control inputs, {i.e. manipulated variables,} shown as yellow dots in the figure, are the mass fill flow, $q_f$, through valve CV1 and the mass vent flow, $q_v$, through valve CV2.
 \item The loop is subject to a disturbance flow, $q_d$, indicated by a red dot in the figure. This disturbance flow is roughly constant at this sampling rate. 
\item Isothermal models are used with temperature as a parameter.
  \end{itemize}
{Accordingly, by \textit{controller inputs} we refer to signals input to the controller, $p_\text{suc}$ and $p_\text{dstl}$; \textit{controller outputs} describe signals output by the controller, $q_f$ and $q_v$.}

\subsection*{Network modeling}
For modeling we proceeded as follows:
\begin{enumerate}[label=(\Roman*)]
\item Individual pipe, branch, valve, compressor, and manifold mass-conserving component models from \cite{SvenRobertBobTCST2021,sven_rob_bob_arXiv} are combined using the \matlab\ method described in \cite{SvenRobertBobTCST2021}, abiding by the Interconnection Rules. This yields a three-input ($q_f$, $q_v$, $q_d$) and two-output ($p_\text{suc}$, $p_\text{dstl}$) continuous-time linear state-space model of dimension 30. This model has the following poles:
\begin{itemize}
\item[--] a single integrator, pole at $s=0$, per Theorem~\ref{thm:integrator1};\label{modeli}
\item[--] all the other poles are stable;
\item[--] some pairs of stable poles have strong oscillatory response, from the resonant behavior, with the natural frequencies of these poles lying above the anti-aliasing filter bandwidth.
\end{itemize}
\item A 46th-order system is constructed by concatenating the continuous-time model~\ref{modeli} with two eighth-order Butterworth low-pass antialiasing filters at 0.4Hz on each of the measured signals, $p_\text{suc}$ and $p_\text{dstl}$;\label{modelii}
\item Balanced truncation is applied to Model~\ref{modelii} to reduce the order to eleven. The integrator mode from Model~\ref{modeli} is unstable and therefore is left invariant by balanced truncation. For model reduction via \texttt{modred} we use the \texttt{Truncate} option to ensure that the strictly proper original model yields a strictly proper reduced model. This, in turn, allows discrete delay-free LQG design with a Kalman filter in place of the default predictor without causing an algebraic loop in \matlab. Obviously, in the real world no algebraic loop would occur. \label{modeliii}
\item The model is discretized at 1 Hz sampling rate using \matlab's function \texttt{c2d}.
\end{enumerate}

\begin{figure}[ht]
\begin{centering}
\includegraphics[width=\columnwidth]{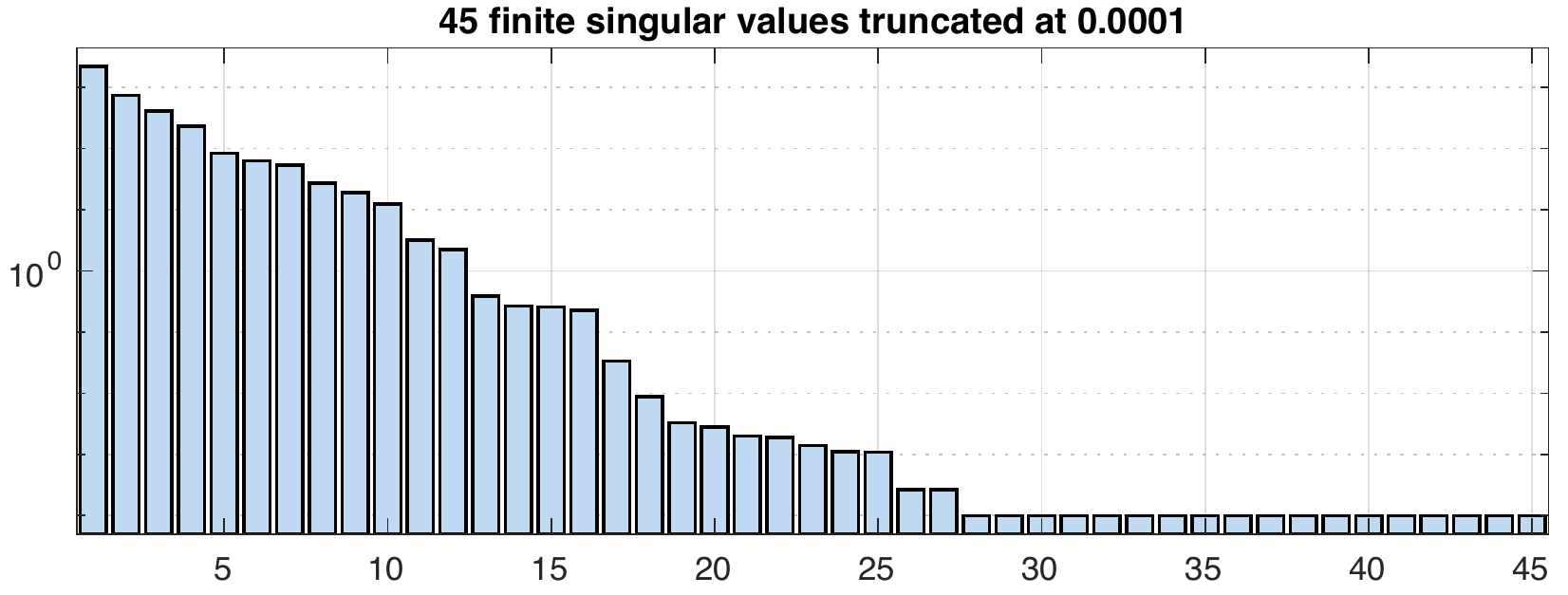}\caption{Hankel singular values of stable part of filtered system. \label{fig:hankel}}
\end{centering}
\end{figure}

Figure~\ref{fig:hankel} shows the Hankel singular values of the stable part of the anti-alias-filtered system. There are 45 values associated with the stable modes plus the integrator, which is excluded from the reduction. For the control design, ten modes were chosen plus the integrator.

\begin{figure}[ht]
    \begin{subfigure}{.45\columnwidth}
\includegraphics[width=\columnwidth]{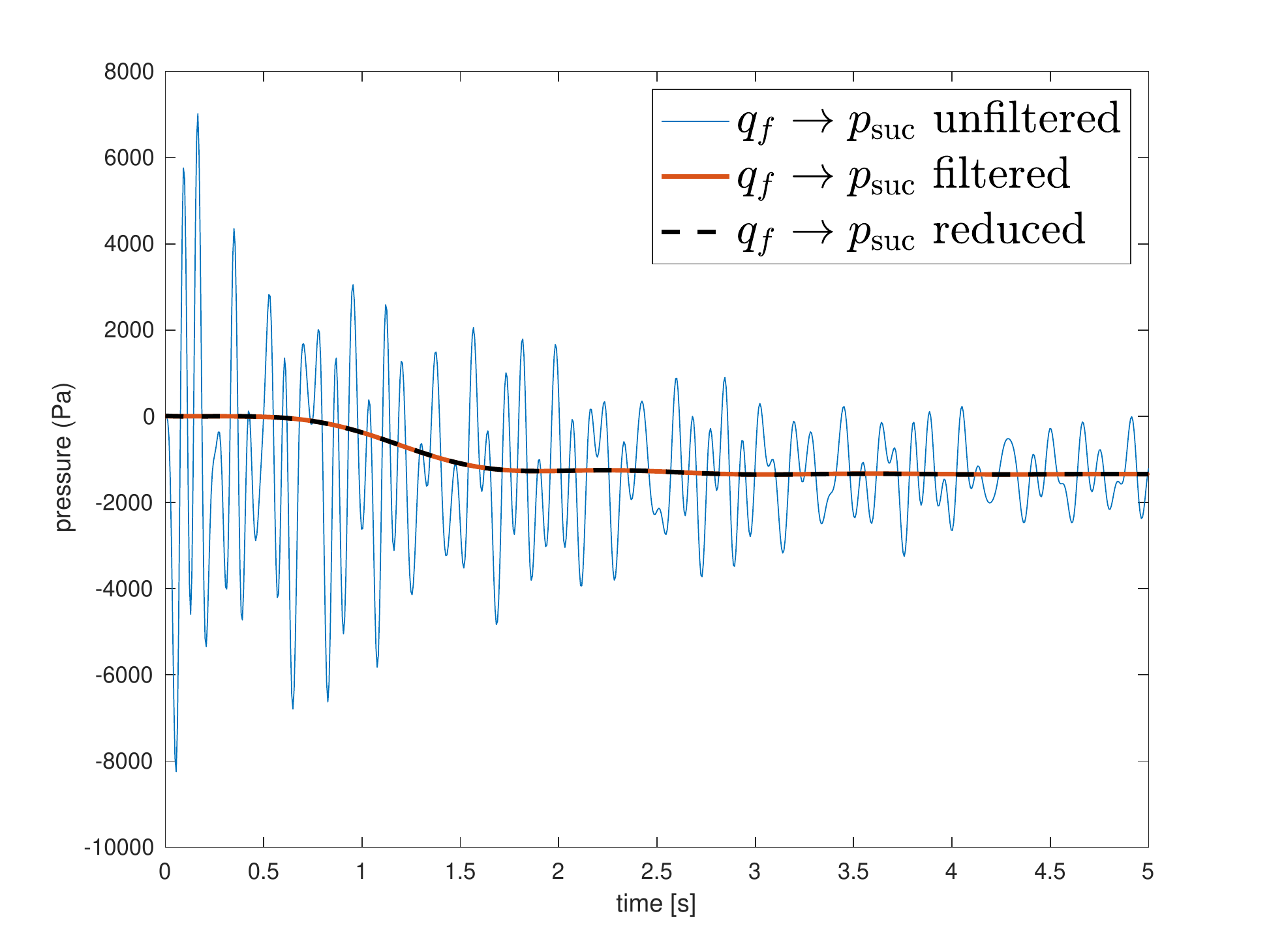}\caption{Fill flow to suction pressure. \label{fig:impulse_fill2suction}}
    \end{subfigure}
    \hfill
    \begin{subfigure}{.45\columnwidth}
\includegraphics[width=\columnwidth]{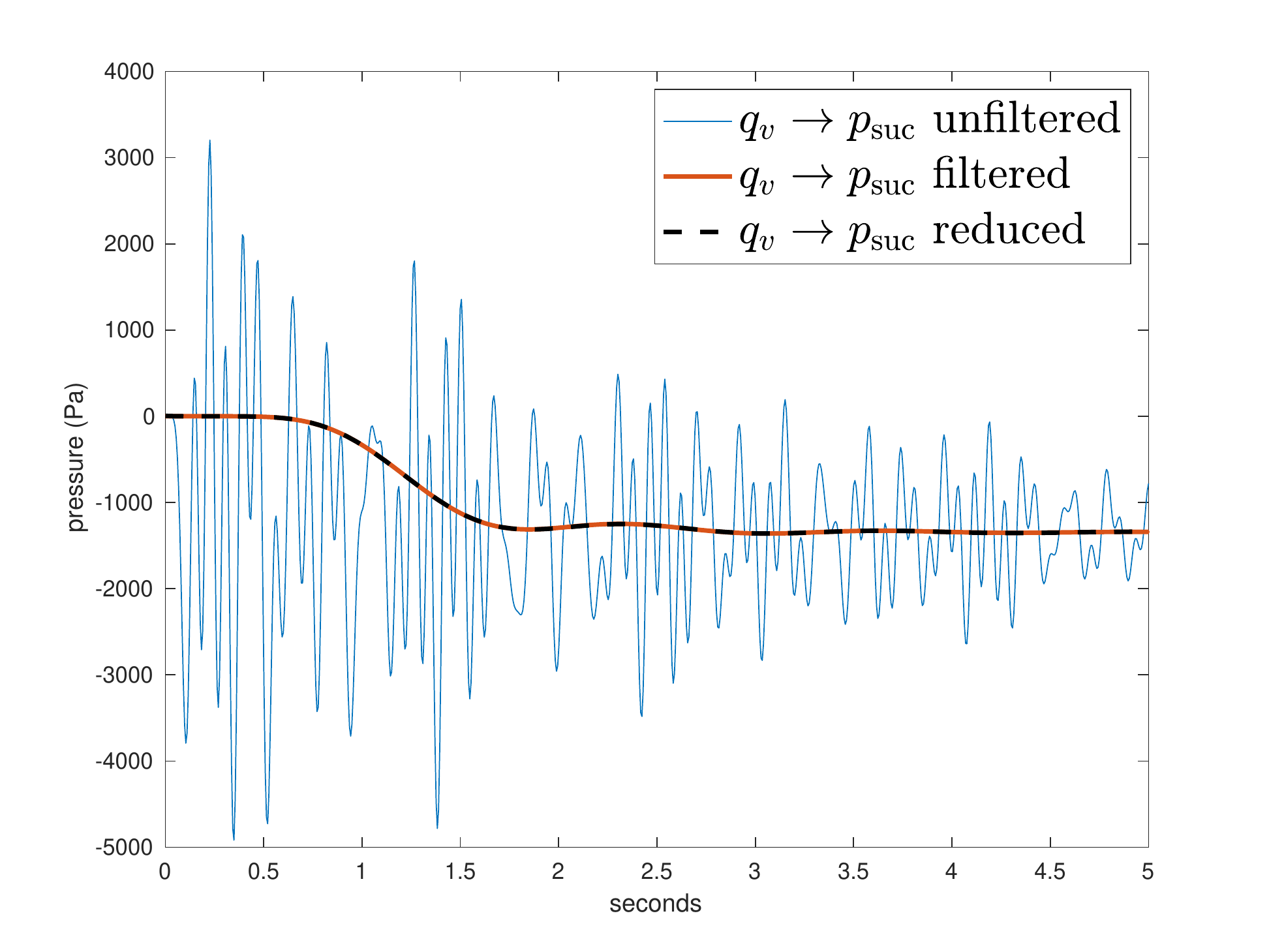}\caption{Vent flow to suction pressure. \label{fig:impulse_vent2suction}}
    \end{subfigure}
    \begin{subfigure}{.45\columnwidth}
\includegraphics[width=\columnwidth]{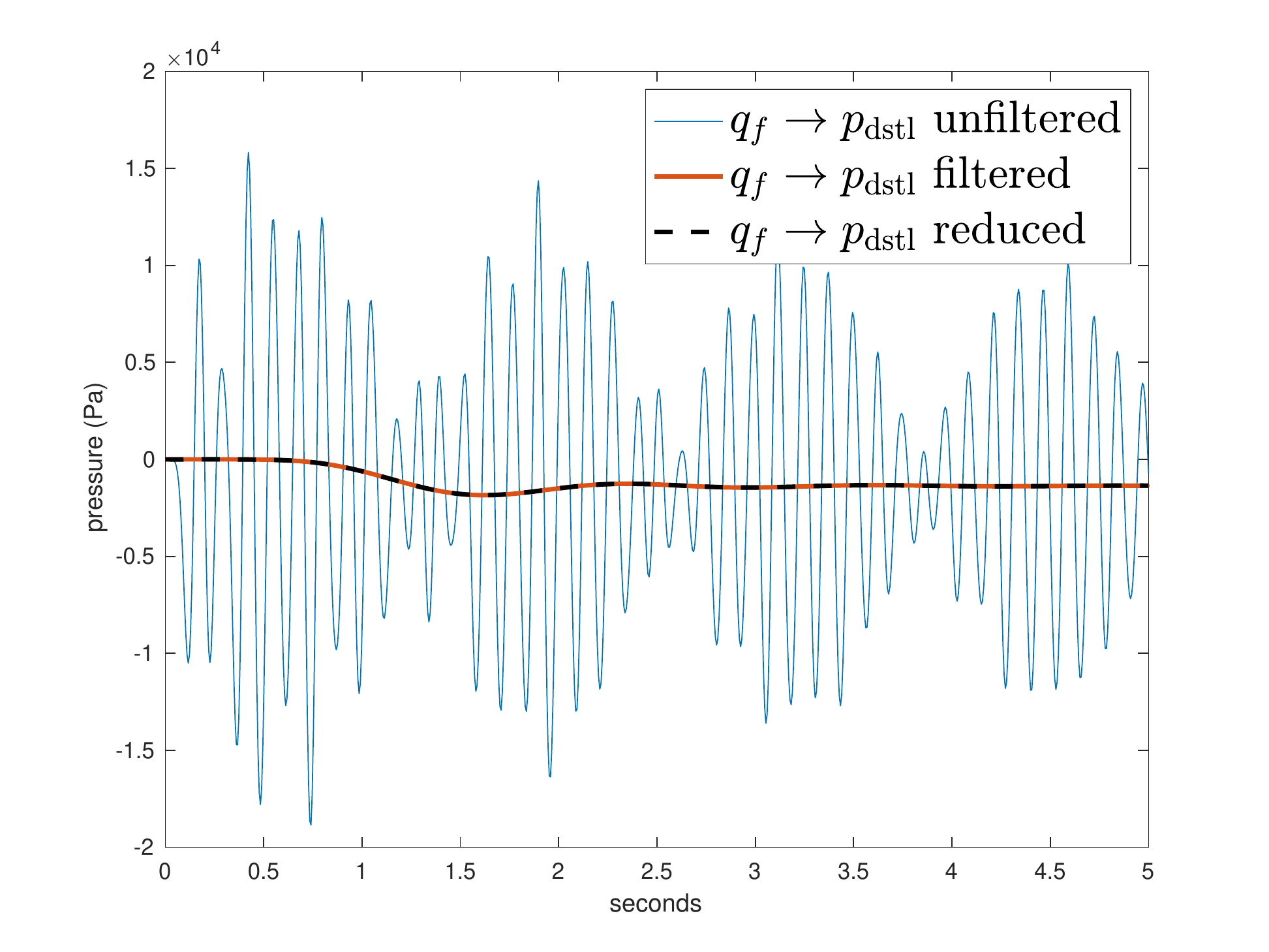}\caption{Fill flow to distal pressure. \label{fig:impulse_fill2distal}}
    \end{subfigure}
    \hfill
    \begin{subfigure}{.45\columnwidth}
\includegraphics[width=\columnwidth]{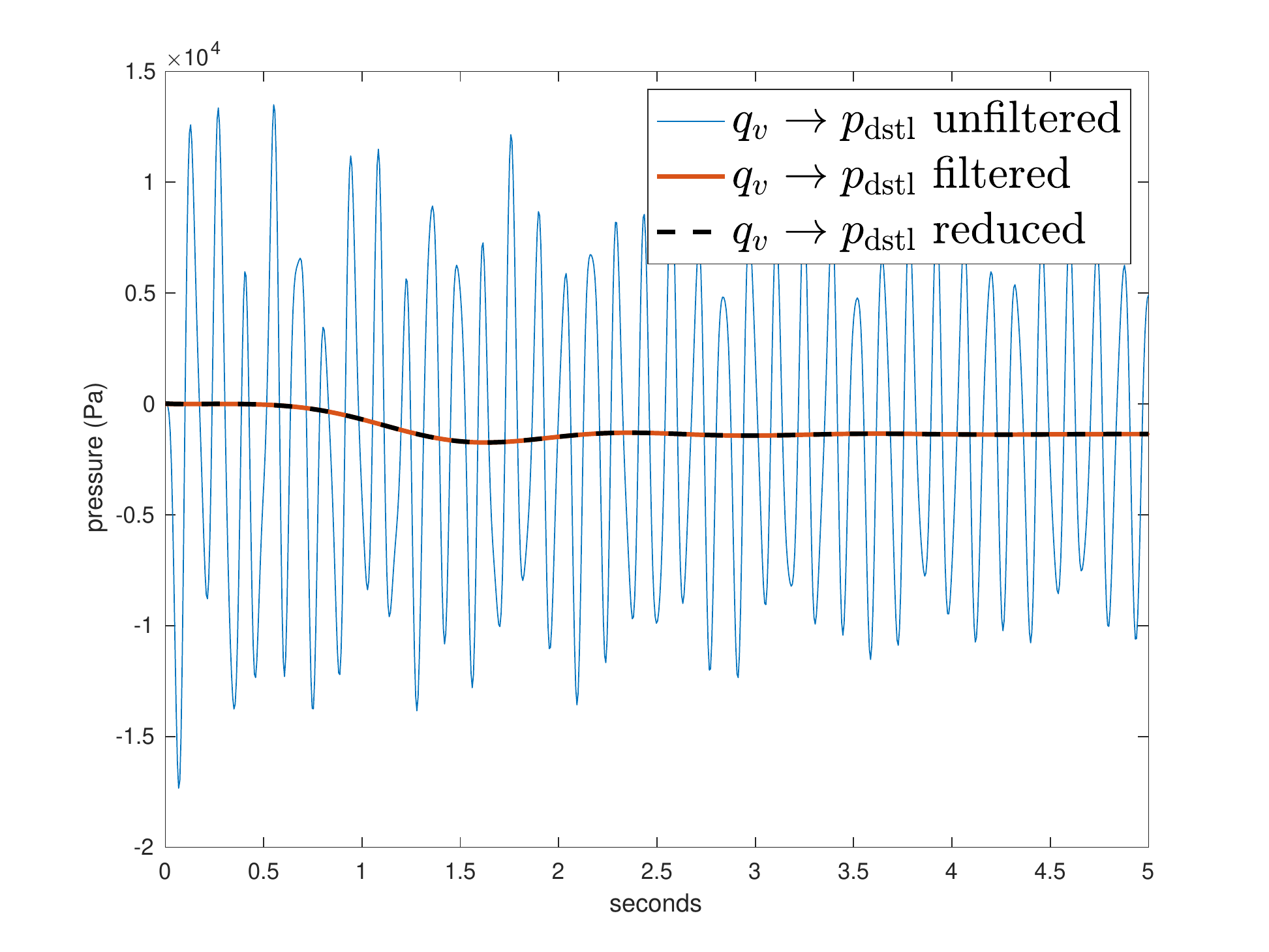}\caption{Vent flow to distal pressure. \label{fig:impulse_vent2distal}}
    \end{subfigure}
    \caption{Loop impulse responses from fill and vent flow to suction and distal pressure for Model~\ref{modeli} (thin blue lines) and anti-aliased Model~\ref{modeliii} (thicker red lines)
    \label{fig:open_loop_impulse}}
\end{figure}

Figure~\ref{fig:open_loop_impulse} displays the four impulse responses of continuous-time Model~\ref{modeli} (thin blue lines) and reduced-order filtered Model~\ref{modeliii} (think red lines) from control inputs to measured outputs.

\subsection*{Open-loop dynamics}
All four plots reveal stable low-frequency behavior and fast oscillations in the unfiltered signals related to the resonant modes of the components comprised by the loop. The non-vanishing offsets show the integrator from mass flow to pressure conservation of mass (Theorem~\ref{thm:integrator1}). The large pressure amplitudes are due to the unit conversion. There is no obvious pairing of any input and output signals, which is problematic for SISO PID controllers assuming decoupled dynamics.

\subsection*{LQG control design}
Since the aim of the project is to facilitate MIMO control design, the standard \matlab\ function \texttt{lqg} is used, firstly without a disturbance model and secondly with integral action, since the control objective is to regulate suction and distal pressures in the face of the step-like disturbance flow.

\begin{figure}[ht]
\begin{centering}
\includegraphics[width=\columnwidth]{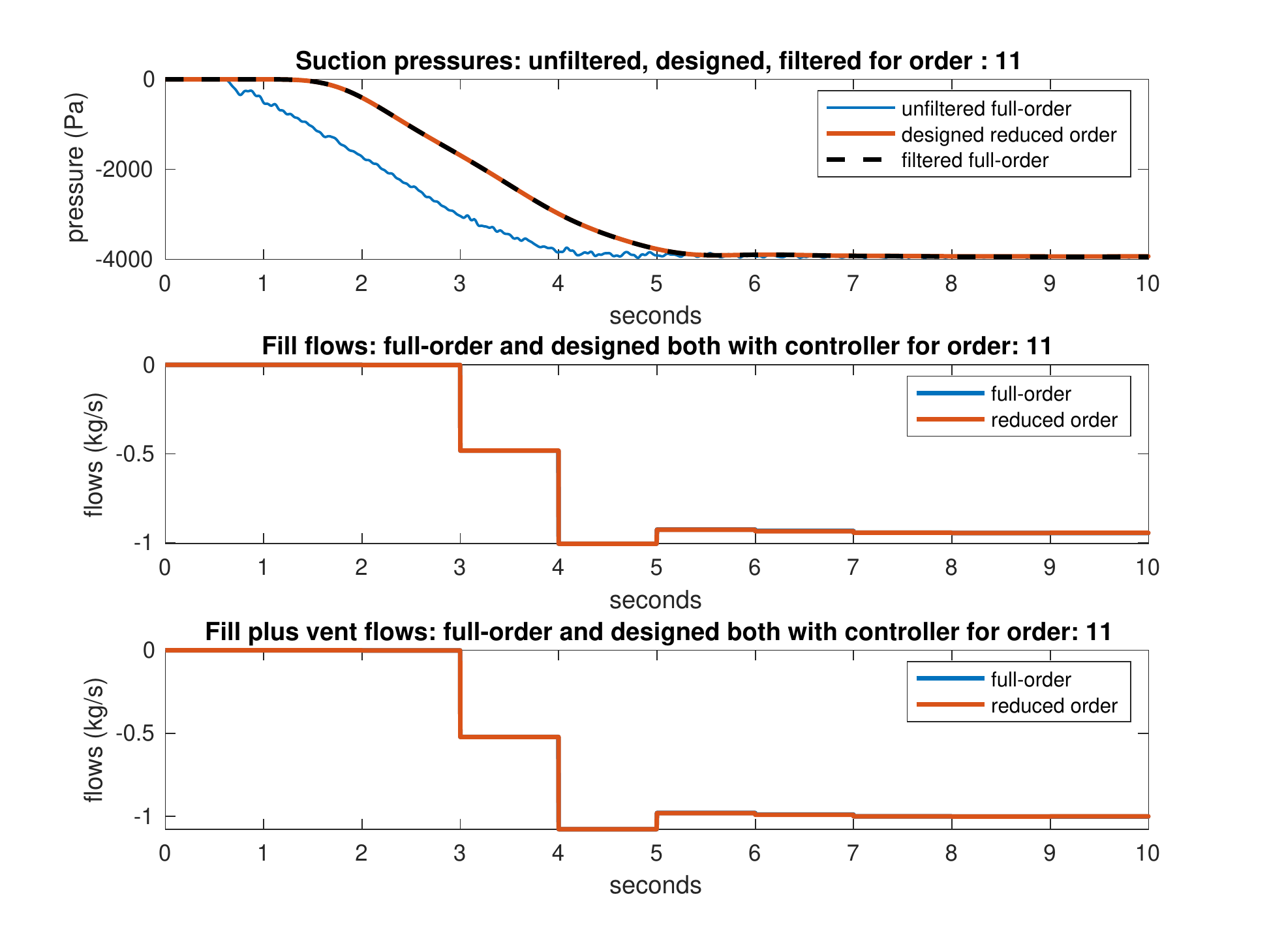}\caption{Disturbance flow to suction pressure closed-loop step responses; step at $0.5$s. \label{fig:distLQG}}
\end{centering}
\end{figure}

Figure~\ref{fig:distLQG} shows closed-loop responses to a step at time $t=0.5$s in the flow out of the loop. The upper plot shows the closed-loop responses of the suction pressures for the three plant models -- the full 30th-order continuous-time network model, the anti-aliased 46th-order model, and the reduced 11th-order model used for the design -- in feedback with the 11th-order LQG digital controller. The center plot shows the corresponding fill flow control input and the lower plot the sum of fill and vent flow inputs. These latter two plots are the outputs of the controller for the closed loops with the full-order plant with antialiasing filters and for the reduced -order plant. Evidently, each closed-loop is stable and there is a steady-state offset to the pressure. The fast unfiltered pressure signal exhibits the underlying resonant system behavior and the absence of anti-aliasing filter group delay. There are very little differences between the anti-aliased full-order responses and the reduced-order.  By imposing larger LQ penalty on the vent flow relative to the fill flow, the controller empahsises the latter.

Per Theorem~\ref{thm:unbound}, since the pressure is finite, the sum of the control inflows exactly matches the disturbance outflow. The LQG design is straightforward and yields static offset of 4kPa (0.6 psi) in the suction pressure for a 1kg/s disturbance flow.

\subsection*{LQG control with integral action}
\begin{figure}[ht]
\begin{centering}
\includegraphics[width=\columnwidth]{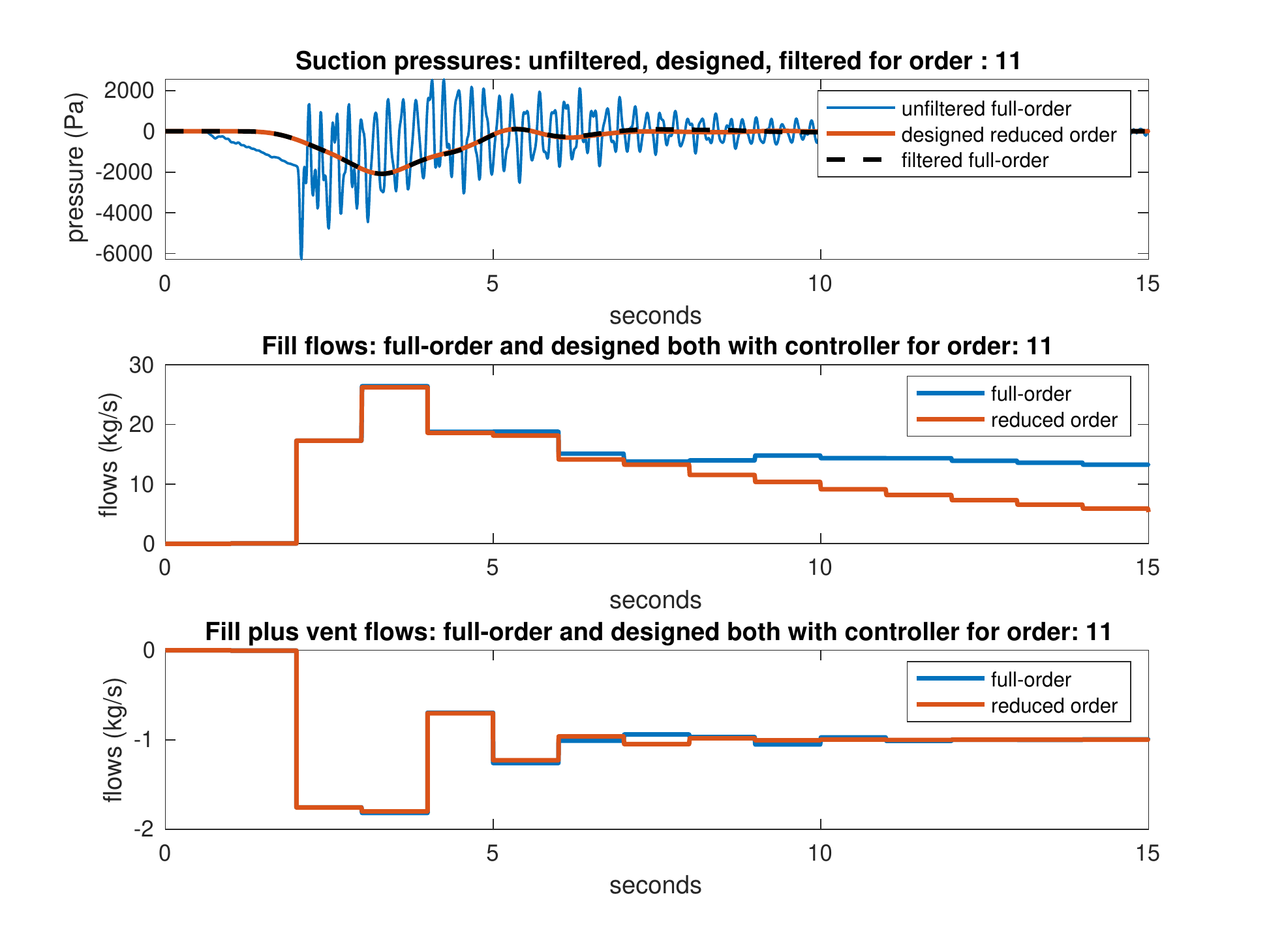}\caption{Disturbance closed-loop step responses with integral action LQG; step at $0.5$s. \label{fig:LQGI}}
\end{centering}
\end{figure}

We next include a modified LQG design with integral action again using \matlab's \texttt{lqg} function, but with this augmentation for the regulation objective. The responses are shown in Figure~\ref{fig:LQGI}.
Several features are evident. 
\begin{enumerate}[label=\arabic*.]
\item Regulation of the suction pressure to zero, i.e. its nominal value, is achieved.
\item Because of the integral action, the closed-loop response is slower than LQG for the same weighting matrices.
\item The 2i2o controller order is now 13 rather than 11 for the LQG case because of the two channels of integral action.
\item The total flows again balance exactly the disturbance in steady state.
\item The regulation objective on both suction and distal pressure forces the vent flow to be applied in spite of its heavier weighting than fill flow.
\item The unfiltered full-order closed-system exhibits resonances and the absence of the anti-aliasing filter group delay.
\end{enumerate}

The conclusion from this example is that the approach of control-oriented modeling and model-based control has led to a MIMO digital controller preserving the well appreciated aspects of LQG control design with and without integral action. Further, the role of the integrators present in the system model and ultimately traceable to conservation of mass led to an easily comprehended control design schema for the simplified models and standard control design tools. The complexity of the composite models and their many resonant modes is suppressed by anti-alias filtering then model reduction. This admits feedback control design focused on managing low frequency bulk behavior unencumbered by acoustic properties of these systems.

\section{Future direction}\label{sec:conclusion}
The paper explores the chain of ideas stemming from replacing DAEs in these fluid models by signal flow graph models more amenable to control design but preserving the algebraic mass conservation properties. These conservation ideas are unraveled in some detail and identified with component transfer function properties. This is shown to be a feature preserved by network interconnections subject to logical connection rules. This, in turn, led to DC gain properties and the presence of integrators, which affect the subsequent control design. This is then examined using a nominal industrial model and standard LQG design packages from \matlab. The important feature of preserving the design aspects is exhibited in this example, thereby validating the claim that these models are \textit{control-oriented,} since the target regulation controller appeared in choices made at each stage of modeling and control design.

While we examined the performance of the designed digital controllers on the resonant continuous-time full-order system, clearly the acid test is to implement and test these controllers on a real plant (a) to ascertain their performance in practice and (b) to appreciate what amenable design flexibility is garnered by MIMO control in this arena.

Extension of these ideas of modeling for control, where the control problem is stated prior to the modeling phase, to other areas, such as thermal systems, minerals processing, materials handling etc would also be of interest. Equally, the methodology explored here might be more thoroughly developed by including ideas from \cite{ObinataAnderson2011} where the control objective plays a central role in modifying the model reduction phase. This permits replacement of balanced truncation by a more control-oriented approach, yielding the modeling, model reduction and control design each to reflect the overarching closed-loop objectives.

\bibliographystyle{ieeetr} 
\bibliography{bobilby} 
\end{document}